\documentclass[twocolumn,10pt,aps,prd,superscriptaddress,nofootinbib]{revtex4-2}

\usepackage{graphicx}
\usepackage{amsmath,amssymb}
\usepackage[dvipsnames]{xcolor}
\usepackage{float}
\usepackage{mathrsfs}
\usepackage{slashed}
\usepackage{braket}
\usepackage{dsfont}
\usepackage{microtype}

\usepackage[colorlinks=true,citecolor=blue,urlcolor=blue]{hyperref}

\begin{document}

\title{Information-Theoretic Gaps in Solar and Reactor Neutrino Oscillation Measurements}

\author{Neetu Raj Singh Chundawat}
\email{chundawat@ihep.ac.cn}
\affiliation{Institute of High Energy Physics, Chinese Academy of Sciences, Beijing 100049, China}
\affiliation{Kaiping Neutrino Research Center, Guangdong 529386, China}

\author{Yu-Feng Li}
\email{liyufeng@ihep.ac.cn}
\affiliation{Institute of High Energy Physics, Chinese Academy of Sciences, Beijing 100049, China}
\affiliation{School of Physical Sciences, University of Chinese Academy of Sciences, Beijing 100049, China}

\begin{abstract}

Quantum estimation theory provides a fundamental framework for analyzing how precisely physical parameters can be estimated from measurements. Neutrino oscillations are characterized by a set of parameters inferred from experiments conducted in different production and detection environments. The two solar oscillation parameters, $\Delta m^2_{21}$ and $\theta_{12}$, can be estimated using both solar neutrino experiments and reactor neutrino experiments. In this letter, we use Quantum Fisher Information (QFI) to quantify and compare the information content accessible in these two experimental setups. For reactor neutrinos, flavor measurements saturate the QFI for both parameters over relevant energy ranges, demonstrating that ideal measurements could achieve ultimate quantum precision bound. In contrast, for solar neutrinos the phase-based QFI vanishes, originating from the quantum coherence, rendering the estimation of $\Delta m_{21}^2$ purely population-based and effectively classical, while the QFI for $\theta_{12}$ remains sizable and nearly saturated by flavor measurements at high energies. Consequently, solar experiments are intrinsically more sensitive to $\theta_{12}$ than to $\Delta m_{21}^2$, reflecting a fundamental difference in the information encoded in the two systems.

\end{abstract}


\maketitle
\newpage

\section{Introduction}
Quantum mechanics is no longer confined to a fundamental theoretical framework, but also underpins a wide range of modern experimental investigations and technologies. It forms the foundation of the ongoing technological revolution. It is now well established in the literature that, in the noiseless scenarios, quantum properties of  systems in consideration can enhance the precision with which parameters governing dynamical processes are estimated. Quantum Estimation Theory (QET), exemplifies how uniquely quantum resources can be exploited to achieve estimation precision beyond classical limits~\cite{Helstrom:1969gj}.

Estimation theory is a central part of statistics that focuses on how information can be extracted most efficiently from data. In many models, some variables are directly measurable observables, while others are latent parameters that characterize the system but cannot be accessed through direct measurement. In many quantum experiments, the measurement process is inherently destructive, so the system cannot be measured repeatedly.  A typical example is a photon being absorbed by a polarization filter, after which it is no longer available for further measurements. In such situations, only the probabilities of the measurement outcomes are relevant, while the post-measurement state plays no role. For systems that are measured only once, generalized measurements described by positive operator-valued measures (POVMs) provide the appropriate framework. Mathematically, a POVM consists of a set of non-negative Hermitian operators acting on the system’s Hilbert space, which are not required to be orthogonal or mutually commuting, and which satisfy a completeness relation~\cite{Paris:2008zgg}.

The parameter estimation problem concerns how effectively collected data, in the form of observables, can be used to infer unknown latent parameters, and what the best achievable precision is for estimating parameters encoded in a quantum state~\cite{Liu:2019xfr, Lu:2010ktl}. By optimizing over all possible POVM measurements, this question is answered by the quantum Cramér-Rao bound (QCRB), which constitutes a central result of QET. The QCRB is determined entirely by the quantum Fisher information (QFI), which depends only on the quantum state. QFI is a central quantity in quantum information theory, with important practical implications in areas such as quantum metrology, where it serves as the quantum analogue of the classical Fisher information~\cite{Giovannetti:2011chh,Albarelli:2020pec,Toth:2014msl}.

Neutrino oscillations~\cite{Pontecorvo:1967fh} provide one of the most powerful probes of physics beyond the Standard Model, and achieving increasingly precise determinations of the oscillation parameters has become a primary objective of current neutrino experiments. In the PMNS parametrization, neutrino oscillations are characterised by three mixing angles $(\theta_{12}, \theta_{13}, \theta_{23})$, two independent mass-squared differences $(\Delta m^2_{21}, \Delta m^2_{31})$, and a CP-violating phase $(\delta_{\text{CP}})$. Among these, solar neutrino experiments predominantly constrain the parameters $\theta_{12}$ and $\Delta m^2_{21}$, which are collectively referred to as the solar oscillation parameters~\cite{SNO:2002tuh,Super-Kamiokande:2008ecj,Borexino:2011ufb}. These oscillation parameters were independently tested and supported by reactor neutrino measurements with very precise determination of solar mass-squared splitting~\cite{KamLAND:2004mhv,KamLAND:2008dgz,JUNO:2025gmd,JUNO:2025fpc}. A coordinated global effort combining data from solar, atmospheric, reactor, and accelerator neutrino experiments has led to progressively more precise determinations of the parameters governing neutrino oscillations~\cite{Esteban:2024eli}.

Solar neutrino experiments and reactor neutrino experiments operate in completely different environments. Solar neutrinos propagate through matter with varying density inside the Sun, then through vacuum, and finally through the matter of the Earth \cite{Mikheyev:1985zog, Davis:1968cp, Wolfenstein:1977ue,Smirnov:2016xzf,Xu:2022wcq,Gann:2021ndb,Maltoni:2015kca}. In reactor neutrino experiments, neutrinos propagate essentially as in vacuum, with matter effects being present but very small \cite{Li:2025hye,Li:2016txk}. 
The agreement between the oscillation parameters $\Delta m^2_{21}$ and $\theta_{12}$ determined from solar neutrino experiments and reactor neutrino experiments had a number of important implications for neutrino physics. More than two decades later, with more measurements, a difference has emerged between the values of $\Delta m^2_{21}$ extracted from solar and reactor neutrino experiments with significance slightly greater than $1.5\sigma$~\cite{Esteban:2024eli}. It may be due to a statistical fluctuation or systematic effects, but it could also turn out to be a manifestation of new physics.

In recent years, there has been growing interest in treating neutrino oscillations as quantum systems within the framework of quantum information \cite{Blasone:2021cau,Blasone:2007vw,Alok:2024jsd,Bouri:2024kcl,Dixit:2018kev,Alok:2025qqr} and in particular QET~\cite{Nogueira:2016qsk,ElBouzaidi:2025mwh,Ignoti:2025rxr,Yadav:2026lsx,Frugiuele:2026yeq}. In this work, we study the QFI associated with reactor and solar neutrino oscillations. Since these two systems exhibit qualitatively different quantum properties, it is essential to analyze them within the framework of QET in order to contrast the information encoded in their respective quantum states and assess the effectiveness of different measurement strategies for parameter estimation. We treat both solar neutrino oscillations and reactor neutrino oscillations within an effective two-flavor framework involving only the solar oscillation parameters.

\section{Fisher Information for Reactor Neutrinos}

In reactor neutrino experiments, neutrinos propagate coherently over macroscopic baselines as in vacuum, and the flavour state remains pure. The evolved state can be written as
\begin{equation}
|\psi(\theta_{12},\phi)\rangle
=
\cos\theta_{12}\,|\nu_1\rangle
+\sin\theta_{12}\,e^{-i2\phi}\,|\nu_2\rangle.
\label{eq:psi_theta_phi}
\end{equation}

We can define the standard oscillation phase accumulated during the neutrino propagation as,
\begin{equation}
\phi \equiv \frac{\Delta m^2_{21}\,L}{4E}
\label{phase}
\end{equation}

where $L$ denotes the baseline and $E$ denotes the neutrino energy. To derive the QFI for these pure state neutrinos, we use Eq.~\eqref{eq:pure_QFI}, given in Appendix~A. For mixing angle, the QFI can be derived as, \(F_Q^{\nu}(\theta_{12})=4\) in the fixed mass basis description and in the fixed flavor basis\footnote{ A unitary transformation which depends on the parameter to be estimated can modify the QFI of the quantum state.} it can be written as \begin{equation}
F_Q^{\nu}(\theta_{12})
=
16\sin^2\phi
\left[
1-\sin^2(2\theta_{12})\cos^2\phi
\right].
\end{equation}
Thus, \(F_Q^{\nu}(\theta_{12})\) represents the maximum amount of Fisher information associated with the mixing angle that can be extracted from the quantum state over all possible quantum measurements. It is evident that the QFI associated with the mixing angle in mass basis is constant and does not depend on the oscillation phase or the detector baseline and in contrast shows this dependency in flavor basis. 

We now compute the Fisher information associated with flavour measurements, since in neutrino oscillation experiments, the oscillation parameters are infered by flavour dependent observables. Such flavour measurements correspond to projective measurements in flavour space, which can be represented as
\begin{equation}
E_e = |\nu_e\rangle\langle\nu_e|,
\qquad
E_\mu = |\nu_\mu\rangle\langle\nu_\mu|,
\qquad
E_e + E_\mu = \mathbb{I}.
\end{equation}

The flavor POVM considered here assumes a binomial measurement of survival~($P_{ee}$) and transition~($1-P_{ee}$) outcomes, whereas reactor neutrino experiments detects surviving electron neutrinos. The probability to detect an electron neutrino is
\(
P_{ee}
= 1 - \sin^2(2\theta_{12})\sin^2\!\left(\phi\right)
\).  Since the measurement has two outcomes, the Fisher information
associated with this POVM is given by
\begin{equation}
F^\nu_{\text{flavor}}(\theta_{12})
=
\frac{\left(\partial_{\theta_{12}} P_{ee}\right)^2}
{P_{ee}\left(1-P_{ee}\right)}
=
\frac{
16\cos^2\!\left(2\theta_{12}\right)\sin^2\!\left(\phi\right)
}{
1 - \sin^2\!\left(2\theta_{12}\right)\sin^2\!\left(\phi\right)
}.
\label{eq:D}
\end{equation}

Hence, from the above expression it is evident that the Fisher information based on flavor POVM satisfies
\(
F_{\text{flavor}}^\nu(\theta_{12}) \le F_Q^{\nu}(\theta_{12})
\),
where $F_Q^{\nu}(\theta_{12})$ is the QFI optimized over all POVMs. The bound is saturated at specific values of the oscillation phase \(\phi\) given in Eq.~\eqref{phase}.

In order to find the QFI for $\Delta m_{21}^2$ we need to use the chain rule for parameters: 
\begin{equation}
F_Q^{\nu}(\Delta m^2_{21}) = F_Q^{\nu}(\phi)\left(\frac{\partial\phi}{\partial(\Delta m^2_{21})}\right)^2.
\label{chain}
\end{equation}

Therefore, using the reparameterization and chain rule, we get the QFI for $\Delta m_{21}^2$ as
\begin{align}
F_Q^{\nu}(\Delta m^2_{21})
&=
\sin^2(2\theta_{12})\left(\frac{L}{2E}\right)^2.
\end{align}

Thus, it is evident from the above expression that the QFI for $\Delta m_{21}^2$ is energy and baseline dependent. Following the same approach as for the mixing angle, we compute the Fisher information for flavour measurements for $\Delta m_{21}^2$, and is given as,  

\begin{equation}
\begin{aligned}
F_{\text{flavor}}^\nu(\Delta m^2_{21})
&=
\frac{\left(\partial_{\Delta m^2_{21}} P_{ee}\right)^2}
{P_{ee}\left(1-P_{ee}\right)}
\\[6pt]
&=
\left(\frac{L}{2E}\right)^2
\frac{\sin^2(2\theta_{12})\,\cos^2\!\left(\phi\right)}
{1-\sin^2(2\theta_{12})\sin^2\!\left(\phi\right)}.
\end{aligned}
\label{eq:Fdm2_flavor}
\end{equation}

The above expression shows that, analogous to the QFI, the Fisher information based on the flavor measurement depends explicitly on both the propagation length and the neutrino energy.

\section{Fisher Information for Solar Neutrinos}For solar neutrinos, we assume adiabatic MSW evolution in the Sun and neglect the contribution of the third mass eigenstate. Due to the long propagation between the Sun and the Earth, the relative phases between the mass eigenstates are effectively averaged out, and the neutrinos arrives as an incoherent mixture, which can be described as,
\begin{equation}
\rho_\odot(\alpha)=p_1(\alpha)\,|\nu_1\rangle\langle\nu_1|
+p_2(\alpha)\,|\nu_2\rangle\langle\nu_2|,
\quad p_1+p_2=1,
\label{eq:rho_solar}
\end{equation}
where $p_1=\cos^2\theta_{12}^m, \, p_2=\sin^2\theta_{12}^m$. The matter-modified mixing angle is given by
\begin{equation}
\cos 2\theta_{12}^m \approx
\frac{\cos 2\theta_{12} - \beta_{12}}
{\sqrt{(\cos 2\theta_{12} - \beta_{12})^2 + \sin^2 2\theta_{12}}},
\label{mix-cos}
\end{equation}
 and the term $\beta_{12}$ is given by
\(
\beta_{12} \equiv \frac{2 V_e^0 E}{\Delta m_{21}^2},
\)
where $V_e^0$ denotes the value of $V_e$ at the
center of the Sun. The survival probability can be written as 
\begin{equation}
P_{ee}^{\odot} \approx \frac{1}{2} + \frac{1}{2}\cos 2\theta_{12}^m \cos 2\theta_{12}.
\label{eq:solprob}
\end{equation}

Solar neutrinos arrive at the detector as an incoherent mixture of mass eigenstates, described by a diagonal density matrix in the mass basis. We use Eq.~\eqref{FI-def} to calculate the QFI associated with solar neutrinos for both solar oscillation parameters and decompose it into two contributions,
\(
F_Q^{\odot}(\alpha) = F_C^{\odot}(\alpha) + F_{\mathrm{rot}}^{\odot}(\alpha)
\). The first term, $F_C^{\odot}$, represents the classical contribution to the QFI and describes the information encoded in
the eigenvalues of the density matrix, while the second term,
$F_{\mathrm{rot}}^{\odot}$, accounts for the parameter dependence
of the eigenvectors. For the mixing angle $\theta_{12}$, the
eigenvectors depend on $\theta_{12}$ through the mass-flavor
basis transformation. Therefore, the basis-rotation term must
be included in the QFI for the mixing angle, whereas it vanishes
for the mass-squared difference $\Delta m_{21}^{2}$. For convenience, we also redefine some of the quantities appearing in the probability expression, as 
\begin{equation}
\delta \equiv \cos 2\theta_{12} - \beta_{12}
\, \text{and} \,
\Delta_m \equiv \sqrt{\delta^2 + \sin^2 2\theta_{12}}.
\label{newdef}
\end{equation}

Using the eigenvalues $\{p_1,p_2\}$, and noting that the
normalization condition implies that all information is encoded in
$p_1$, it is sufficient to compute the Fisher information for this single eigenvalue. We obtain the final expression for population-based term of QFI for mixing angle in solar neutrino case,
\begin{equation}
F_C^{\odot}(\theta_{12})
=
\frac{4\left(\sin^22\theta_{12}
+\delta\cos2\theta_{12}\right)^2}{\Delta_m^{\,4}}.
\label{eq:qfi_theta_final1}
\end{equation}
For a two level system, we can write $F_{rot}^{\odot}(\theta_{12}) = 4(p_1-p_2)^2$, and for solar case, we obtain it to be,
\begin{equation}
F_{rot}^{\odot}(\theta_{12})
=
4\left(\frac{\delta}{\Delta_m}\right)^2 .
\label{eq:qfi_theta_final2}
\end{equation}

\begin{figure*}[htb]
    \centering
    \includegraphics[width=1\linewidth]{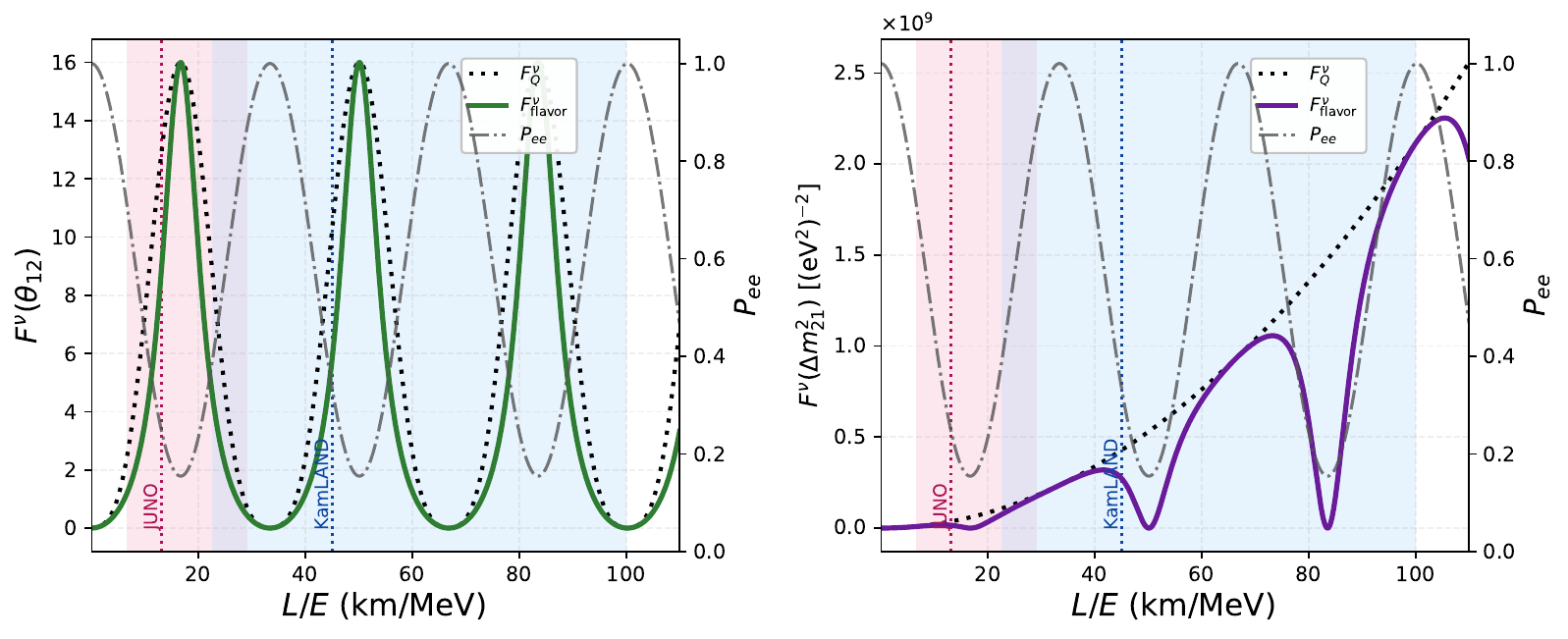}
    \caption{The left panel shows the Fisher information for $\theta_{12}$, while the right panel shows the Fisher information for $\Delta m_{21}^{2}$ as a function of $L/E$. In both panels, the black dotted curve represents the intrinsic QFI, whereas the colored solid curves denote the flavor Fisher information for $\theta_{12}$ (green) and $\Delta m_{21}^{2}$ (purple). The dot-dashed curve, shown on the right vertical axis, represents the electron neutrino survival probability. The light red and blue shaded regions indicate the approximate $L/E$ ranges covered by JUNO and KamLAND, respectively, for neutrino energies in the range $E=1.8$-$8~\mathrm{MeV}$. The vertical dashed lines correspond to the peak reactor antineutrino energy, $E=4~\mathrm{MeV}$, for the two reactor baselines.}
    \label{fig:FIvsLE}
\end{figure*}

To determine the Fisher information for the flavour measurement schemes, we use the solar survival probability given in Eq.~\eqref{eq:solprob}. Thus, it is obtained for the flavour POVM as
\begin{equation}
F_{\text{flavor}}^{\odot}(\theta_{12})
=
\frac{4\left(\delta^3+\delta+\cos2\theta_{12}\sin^22\theta_{12}\right)^2}{(1+\delta^2)\,\Delta_m^{\,4}}.
\label{flavor_FI_final}
\end{equation}

In the case of mass-squared difference, it is sufficient to evaluate only the population contribution $F_C^{\odot}(\Delta m_{21}^2)$. Using the chain rule, we obtain the final expression as,
\begin{equation}
F_C^{\odot}(\Delta m^2_{21})
=
\left(\frac{\beta_{12}}{\Delta m^2_{21}}\right)^2
\frac{\sin^22\theta_{12}}{\Delta_m^{\,4}}.
\label{eq:qfi_dm2}
\end{equation}

We obtain the Fisher information for the flavour POVM associated with $\Delta m^2_{21}$ as
\begin{equation}
F_{\text{flavor}}^{\odot}(\Delta m^2_{21})
=
\left(\frac{\beta_{12}}{\Delta m^2_{21}}\right)^2
\frac{\cos^22\theta_{12}\,\sin^22\theta_{12}}{(1+\delta^2)\,\Delta_m^{\,4}}.
\end{equation}

\section{Results and Discussions} For the case of reactor neutrino experiments, we take the baselines of JUNO ($52.5\,\mathrm{km}$) and KamLAND ($180\,\mathrm{km}$), together with the most recent values of the relevant oscillation parameters~\cite{Esteban:2024eli}, and show the QFI and flavor Fisher information as functions of $L/E$ in Fig.~\ref{fig:FIvsLE}.  We also display the electron neutrino survival probability $P_{ee}$ and indicate the $L/E$ regions relevant for JUNO and KamLAND by the red and blue shaded bands, respectively. 

It is evident from the figure that for the solar mixing angle, both the QFI and the flavor Fisher information exhibits an oscillatory structure with maxima where it saturates the quantum limit of information encoded in the neutrino state. Since the mixing angle $\theta_{12}$ primarily governs the oscillation amplitude, the survival probability becomes insensitive to $\theta_{12}$ at its maxima, causing the flavor Fisher information to vanish. In contrast, at the minima of $P_{ee}$, variations in $\theta_{12}$ produce the largest change in the oscillation amplitude, leading to maximal flavor Fisher information saturating the QFI bound.  

For the mass-squared difference, the QFI increases monotonically with increasing $L/E$, whereas the flavor Fisher information exhibits an oscillatory structure with alternating maxima and minima. The scale of order $10^{9}$ arises from the choice of parameterization and the corresponding units of $\Delta m_{21}^{2}$, and therefore its magnitude cannot be directly compared with that of mixing angle. The derivative of $P_{ee}$ at its minima with respect to $\Delta m_{21}^2$ goes to zero while the denominator of Eq.~\eqref{eq:Fdm2_flavor} remains nonzero, causing the flavor Fisher information to vanish. However, at the maxima of $P_{ee}$, both numerator and denominator vanishes but the ratio reaches finite limit, making flavor Fisher information to approach QFI. This complementary behavior demonstrates that the information for estimating $\theta_{12}$ and $\Delta m_{21}^{2}$ is distributed differently over the oscillation cycle. Therefore, no particular value of $L/E$ is simultaneously optimal for estimating both parameters through flavor measurements, highlighting the importance of broad spectral measurements.

\begin{figure}[htb!]
    \centering
    \includegraphics[width=1\linewidth]{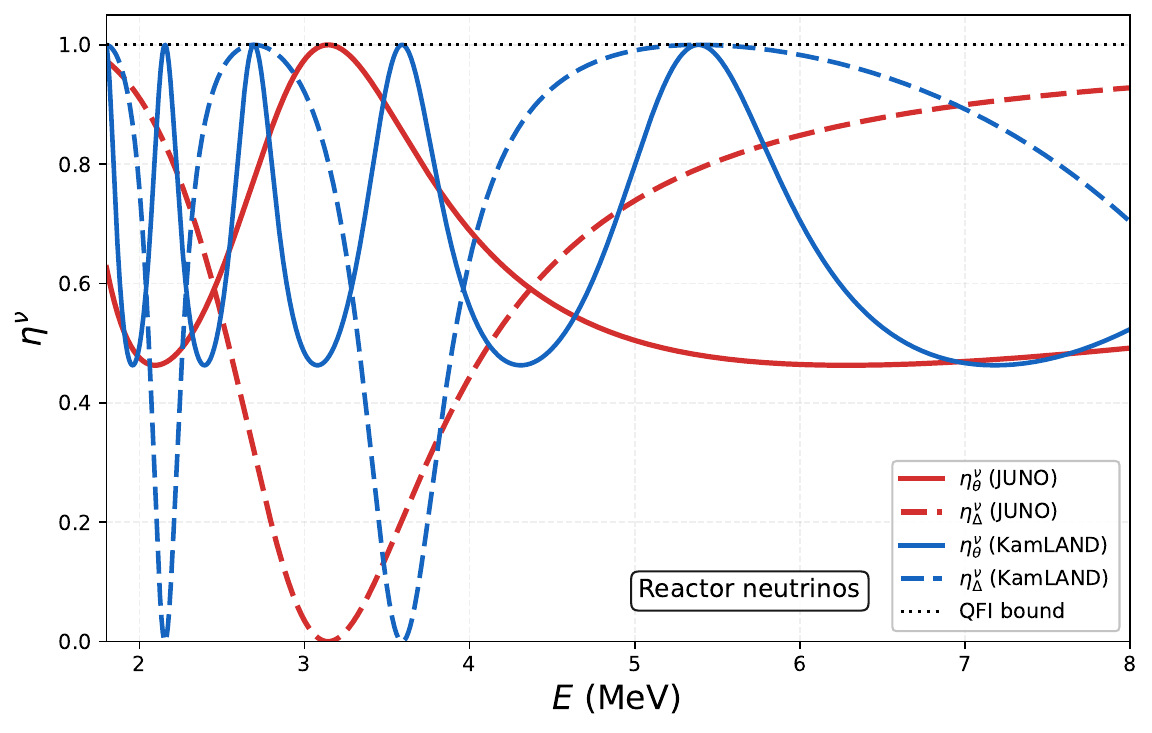}
    \caption{The extracted fraction \(\eta^{\nu}\) for reactor neutrinos as a function of the neutrino energy \(E\) for the JUNO and KamLAND baselines. The red curves correspond to JUNO, while the blue curves correspond to KamLAND. For each baseline, the solid curve represents \(\eta^{\nu}_{\theta}\), and the dashed curve represents \(\eta^{\nu}_{\Delta}\), associated with \(\theta_{12}\) and
\(\Delta m^2_{21}\), respectively. The black dotted line denotes the quantum limit set by the QFI.
}
    \label{fig:reactor}
\end{figure}

We also plot the quantity $\eta^{\nu}$, defined as the ratio of the Fisher information extracted using a flavor POVM to the QFI available in the state. Here, the ratio $\eta^{\nu}$ is defined for electron-flavor POVM only. For the mixing angle $\theta_{12}$ as can be seen from Fig.~\ref{fig:reactor}, the ratio $\eta^{\nu}_\theta$ for KamLAND reaches the QFI bound at certain energies below $6\,\mathrm{MeV}$ and remains finite over the full energy range. For most of its characteristic energy range, the information extraction efficiency for $\theta_{12}$ remains large, above approximately $0.5$.  It is to be noted that at the maxima of $\eta^{\nu}_\theta$, both QFI and flavor Fisher information vanishes, so the efficiency is formally undefined, although its limiting value reaches unity. For JUNO, the extracted fraction $\eta^{\nu}_{\theta}$ follows a behaviour similar to that observed for KamLAND and reaches large values around $E\simeq 3,\mathrm{MeV}$, exhibiting a broader peak compared to KamLAND. This energy coincides with the mean reactor antineutrino energy and shows that flavour measurements in this region have access to maximum information available on the parameter $\theta_{12}$. At higher energies, $\eta^{\nu}_{\theta}$ decreases, indicating that JUNO is particularly optimal for estimating $\theta_{12}$ in the lower-energy region of the reactor neutrino spectrum.

 \begin{figure*} 
        \centering
        \includegraphics[width=1\linewidth]{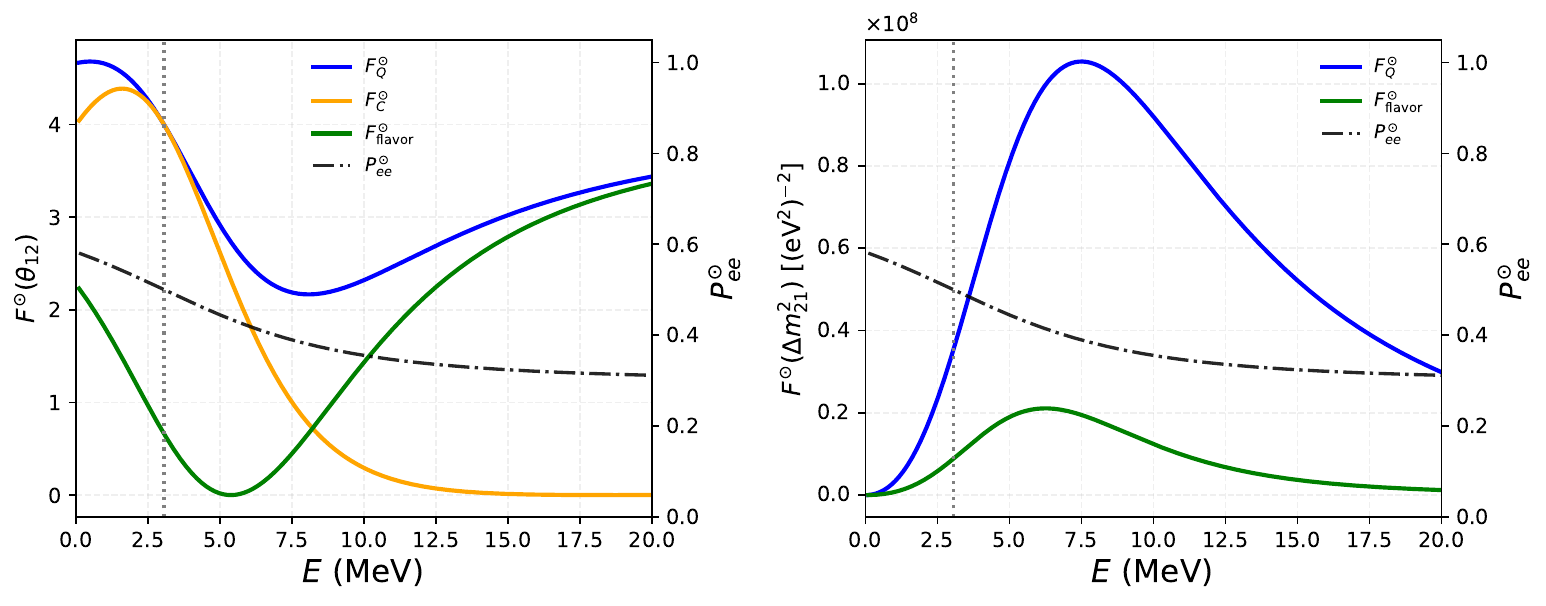}
        \caption{The left panel shows the Fisher information for $\theta_{12}$, while the right panel corresponds to $\Delta m_{21}^2$ as a function of neutrino energy $E$ for solar neutrinos. In the left panel, the blue curve denotes the QFI, the orange curve represents the population contribution, and the green curve shows the Fisher information associated with the flavor POVM. In the right panel, the blue and green curves correspond to the QFI and the flavor POVM Fisher information for $\Delta m_{21}^2$, respectively. The electron-neutrino survival probability $P_{ee}^\odot$ is indicated by the black dot-dashed curve, and the MSW resonance point is marked by the black dotted line. 
}
        \label{fig:solarvsE}
    \end{figure*}

For the mass-squared splitting $\Delta m_{21}^2$, the behaviour of $\eta^{\nu}_\Delta$ is similar to that of $\eta^{\nu}_\theta$ below $4\,\mathrm{MeV}$ for KamLAND. However, at higher energies the ratio $\eta^{\nu}_\Delta$ remains saturated over a wider energy range, indicating that the flavor POVM continues to extract a significant fraction of the available QFI. This suggests that flavor measurements constitute an effective measurement strategy for estimating $\Delta m_{21}^2$ for KamLAND.  In the case of JUNO, the ratio $\eta^{\nu}_\Delta$ exhibits a nontrivial energy dependence. Near the inverse beta decay threshold, $\eta^{\nu}_{\Delta}$ starts close to unity and decreases to nearly zero around $E\simeq 3\,\mathrm{MeV}$. Beyond this energy, $\eta^{\nu}_{\Delta}$ increases again but does not reach the QFI bound up to $8\,\mathrm{MeV}$. Nevertheless, for $E\gtrsim 4\,\mathrm{MeV}$ the extracted fraction remains larger than about $0.5$, indicating that a substantial amount of Fisher information is accessible over this energy range. This implies that flavor measurements even not fully optimal at higher energies, they still enable a precise determination of $\Delta m_{21}^2$ due to the large fraction of information extracted.

Since solar neutrinos constitute a qualitatively different quantum system from an information-theoretic standpoint, we analyze the same set of quantities as in the reactor case using the eigenvalue structure of the density matrix. We organize the discussion into three distinct energy regimes:

\begin{itemize}
    \item \textit{Low-energy Region:\\} In the low-energy limit ($E<1 \, \mathrm {MeV}$), where $\beta_{12}\ll 1$ , neutrino propagation is vacuum-like and matter effects can be neglected, so that $\theta_{12}^m \to \theta_{12}$. In this regime, the QFI for $\theta_{12}$ attains its maximum value over the entire energy range, as shown by the blue curve in Fig.~\ref{fig:solarvsE}. Using the corresponding low-energy approximation in Eq.~\eqref{eq:qfi_theta_final1} and \eqref{eq:qfi_theta_final2}, the QFI is found to be close to $4$ and decreases monotonically as $\beta_{12}$ increases. The population contribution, $F_C^\odot(\theta_{12})$, indicated by the orange curve, closely follows the total QFI and therefore dominates the information content in this regime. The dot-dashed black curve shows the electron neutrino survival probability $P_{ee}^\odot$ for reference. The Fisher information associated with the flavor POVM is represented by the green curve. Since in this limit $P_{ee}^\odot$ depends only on $\theta_{12}$, flavor measurements extract a considerable amount of information on this parameter from the total QFI available in the system. Since the extracted information remains well below the quantum limit, flavor measurements constitute a suboptimal measurement strategy in the low-energy regime.  In the case of $\Delta m^2_{21}$, basis-rotation contribution vanishes, and the available information arises solely from the population term and the QFI ($F_Q^{\odot}(\Delta m^2_{21}) \simeq F_C^{\odot}(\Delta m^2_{21})$) is negligible at low neutrino energies. As the energy increases, the $F_Q^{\odot}(\Delta m^2_{21})$ in the solar neutrino state begins to rise, as shown by the blue curve in Fig.~\ref{fig:solarvsE}. For flavor measurements, the solar survival probability $P_{ee}^{\odot}$ depends primarily on the mixing angle, and variations in $\Delta m^2_{21}$ have little impact on them at low energies. Consequently, the Fisher information associated with the flavor POVM also vanishes in this limit and increases only as the energy grows and the survival probability departs from its vacuum value. In this region, flavor measurements extract a non-negligible fraction of the information on $\Delta m^2_{21}$, however, the total information available in the solar neutrino state remains intrinsically limited.

    \item \textit{MSW Transition Region:}\\
    In this energy range, $\theta^m_{12}$ varies rapidly with energy, approximately over $E\simeq 1$-$6$~MeV. As a result, the QFI for $\theta_{12}$ decreases sharply from its low-energy value, as shown by the blue curve in Fig.~\ref{fig:solarvsE}, with the population contribution (orange curve) exhibiting the same behavior and providing the dominant contribution. This indicates that the MSW effect does not enhance the information on $\theta_{12}$; instead, the precision with which the mixing angle can be estimated is reduced in the transition region compared to the vacuum-dominated regime. For the flavor POVM, the Fisher information (green curve) follows a similar trend but drops to zero in this energy interval, demonstrating that flavor measurements become strongly suboptimal for the determination of $\theta_{12}$ within the MSW transition region. In contrast to $\theta_{12}$, the QFI for $\Delta m_{21}^2$ increases in the MSW transition region, since the location of the transition is governed by the mass-squared difference. Consequently, the available information on $\Delta m_{21}^2$ is encoded in the energy dependence of the MSW turn-over, corresponding to spectral Fisher information rather than oscillatory phase information. Similarly, the Fisher information associated with the flavor POVM increases in this energy range. However, as shown in the right panel of Fig.~\ref{fig:solarvsE}, it does not saturate the total QFI of the state. This indicates that, although flavor measurements become more informative for $\Delta m_{21}^2$ in the MSW region, they remain a suboptimal measurement strategy, with a substantial gap persisting between the accessible information and the total information available in the state.
    
\item \textit{High-Energy Region:}\\
In the matter-dominated regime $(E> 10 \,\mathrm {MeV})$, where $\beta_{12}\gg 1$, the QFI for $\theta_{12}$ increases again. Towards the high-energy end of the spectrum, as $P_{ee}^\odot$ becomes solely a function of $\theta_{12}$, the corresponding QFI approaches values of order $3$-$4$. In this region, the population contribution to the QFI becomes negligible for $\theta_{12}$, while the basis-rotation contribution given in Eq.~\eqref{eq:qfi_theta_final2} dominates the total information. The Fisher information associated with the flavor POVM closely follows the QFI and, unlike in the low-energy regime, approaches the maximum information available in the state. Consequently, flavor measurements become effectively optimal in the matter-dominated region, nearly saturating the quantum limit. For the mass-squared difference, the QFI decreases at high energies, with the Fisher information associated with the flavor POVM exhibiting a similar trend. Since $P_{ee}^\odot$ becomes independent of $\Delta m_{21}^2$ and depends only on the mixing angle in this regime, the corresponding flavor Fisher information vanishes in the high-energy limit. This demonstrates that flavor measurements are strongly suboptimal for estimating $\Delta m_{21}^2$ in the matter-dominated region and are unable to access most of the information available in the quantum state.
\end{itemize}

\begin{figure}[htb]
        \centering
        \includegraphics[width=1\linewidth]{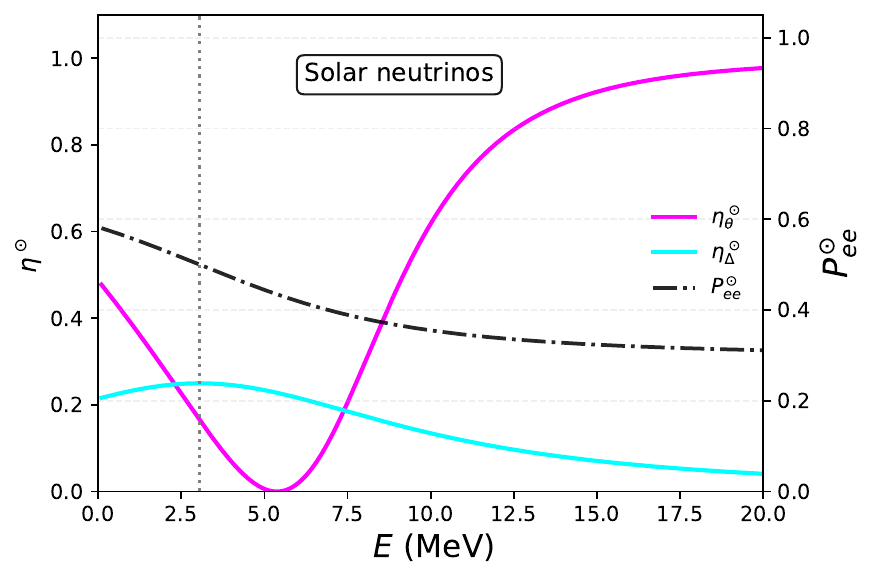}
        \caption{Extracted fraction $\eta^{\odot}$ as a function of the neutrino energy $E$ for solar neutrinos. The magenta and cyan curves correspond to $\eta^{\odot}$ for $\theta_{12}$ and $\Delta m_{21}^2$, respectively.}
        \label{fig:eta_solar}
    \end{figure}

We define, in analogy with the reactor case, the extracted fraction $\eta^\odot$ for both oscillation parameters, which quantifies the fraction of the total quantum Fisher information accessible through electron-flavor measurements, as shown in Fig.~\ref{fig:eta_solar}. For $\theta_{12}$, $\eta^\odot_{\theta}$ is sizable at low energies, decreases in the MSW transition region, and rises again in the matter-dominated regime, approaching unity at high energies, indicating that flavor measurements nearly saturate the quantum limit and provide a robust determination of $\theta_{12}$. In contrast, for $\Delta m_{21}^2$, the extracted fraction $\eta^\odot_{\Delta}$ is of order $0.2$ at low energies, increases moderately and attains a maximum near the MSW resonance, reflecting the sensitivity of flavor observables to shifts in the resonance position, and then decreases significantly at higher energies while remaining small but nonvanishing toward the high-energy end of the spectrum.

In this analysis, we use the two-flavor approximation for both reactor and solar neutrino oscillations, focusing only on the solar oscillation parameters. The three-flavor corrections arising from the relatively small value of $\theta_{13}$ are neglected. These corrections to the oscillation probabilities are expected to produce only minor quantitative changes in the values of the QFI and the corresponding flavor Fisher information, while the main features remain qualitatively unchanged in the single parameter estimation scenario. An extension of the present analysis to the full three-flavor framework, including all oscillation parameters and their mutual correlations, is required to obtain a complete information-theoretic understanding of neutrino oscillation measurements within the framework of multiparameter QET~\cite{Huang:2026bws}.

\begin{table*}[htb]
\centering
\caption{Illustrative relative precision obtained from the QFI and flavor
Fisher information at fixed neutrino energy.}
\label{tab:fixed_energy_demo}
\begin{tabular}{lccccc}
\hline\hline
& & \multicolumn{2}{c}{$\sin^2\theta_{12}$}
& \multicolumn{2}{c}{$\Delta m_{21}^2$} \\
\cline{3-4}\cline{5-6}
Case & $N$ & QFI & Flavor POVM  & QFI & Flavor POVM  \\
\hline
JUNO-like     & $10^4$ & $0.85\%$ & $1.00\%$ & $0.44\%$ & $0.69\%$ \\
KamLAND-like  & $10^2$ & $9.2\%$  & $11.7\%$ & $1.28\%$ & $1.70\%$ \\
Solar-like         & $10^4$ & $2.02\%$  & $2.57\%$  & $1.39\%$  & $4.66\%$ \\
\hline\hline
\end{tabular}
\end{table*}
\section{Fisher information and Precision bounds}

The QFI and Fisher information associated with the flavor POVM discussed in this work quantify the sensitivity of the neutrino state and the corresponding probability distribution to changes in the oscillation parameters for a single neutrino. In neutrino experiments, a large number of neutrinos are observed, each contributing information about the oscillation parameters. Therefore, estimating the achievable precision from these Fisher information quantities also requires considering the number of independent measurements. As follows from the Cramér-Rao bound (CRB) in Eq.~\eqref{eq:CRB_classical}, a sufficiently large number of measurements $N$ can lead to a smaller uncertainty bound and hence better precision. Thus, it is important to consider $N$ for both reactor and solar neutrino experiments and compare the precision bounds obtained from the quantum limit and the flavor measurement scheme. 

To illustrate the role of statistics, we consider representative
experimental configurations at fixed neutrino energies. For the
JUNO- and KamLAND-like configurations, we take $N=10^4$ and
$N=10^2$, respectively, at a neutrino energy of $4~\mathrm{MeV}$.
For the solar case, we assume the same statistics as JUNO,
with a neutrino energy of $10~\mathrm{MeV}$ and an effective
matter potential of $V_e^0=5.0\times10^{-12}~\mathrm{eV}$.
We evaluate the relative uncertainities, given by corresponding CRBs,
using the Fisher information quantities derived in the
preceding sections. The relative precision in the
solar oscillation parameters for all three example configurations are presented in Table~\ref{tab:fixed_energy_demo}.

As can be seen from Table~\ref{tab:fixed_energy_demo}, the JUNO-like configuration can achieve better precision than the KamLAND-like configuration due to its larger $N$, although the latter has higher single-neutrino Fisher information for $\Delta m_{21}^{2}$ at $4~\mathrm{MeV}$. For both reactor configurations, the flavor POVM precision bounds remain relatively close to the QFI bounds, indicating that flavor measurements can provide a near-optimal measurement strategy for both $\theta_{12}$ and $\Delta m_{21}^{2}$. On the other hand, considering the same $N$ for the JUNO-like and solar-like configurations, the solar-like configuration exhibits a larger loss of information for $\Delta m_{21}^{2}$ than for $\theta_{12}$ when going from the quantum state to the measurement level. This suggests that flavor measurements are relatively close to optimal for $\theta_{12}$ but suboptimal for $\Delta m_{21}^{2}$ in the solar case. In a more comprehensive analysis, the fixed-energy approximation will be replaced by reactor event spectra, while the present estimates serve only to illustrate the precision bounds obtained by including information from both the survival and transition channels.

Furthermore, all the results presented in this work are obtained assuming perfect flavor measurements, corresponding to an ideal flavor POVM. In neutrino experiments, however, detector-specific effects also play an important role in determining the experimentally accessible Fisher information. In Ref.~\cite{Chundawat:2026lcm}, the authors demonstrates a quantitative comparison between the intrinsic QFI and the Fisher information computed from reconstructed event spectra using Poisson statistics. Such an approach incorporates detector effects such as finite event statistics, energy resolution, backgrounds, and systematic uncertainties, thereby enabling a comparison between the information predicted by the QET framework and the experimentally accessible information.

\section{Final Remarks} We examined parameter estimation in neutrino oscillation experiments from an information-theoretic perspective. Solar and reactor neutrino experiments both provide precise determinations of the relevant oscillation parameters, albeit with markedly different levels of precision. We address this disparity at a fundamental level by asking whether it is governed by the amount of information encoded in the quantum state at the time of detection for the parameter under consideration. We show that, for neutrinos produced and propagated under different physical conditions, such as in reactor and solar environments, the amount of Fisher information available for estimating the same parameter can differ, leading to different achievable precisions. Our analysis shows that even for the same neutrino state, the information content can differ substantially between different parameters, leading to distinct estimation capabilities. As in the solar neutrino case, flavor measurements are suboptimal for the determination of $\Delta m_{21}^2$ but remain effective for estimating $\theta_{12}$. In contrast, for reactor neutrino experiments, flavor measurements are optimal for the determination of both oscillation parameters.

\textit{\textbf{Acknowledgements.}}---
NRSC would like to thank Luis A.~Delgadillo for useful discussions. This work was supported in part by the National Natural Science Foundation of
China under grant number 12075255. We thank the anonymous referee for their comments, which have helped us improve the completeness of the discussion.

\appendix
\label{FI}
\section{Quantum Fisher Information}
\renewcommand{\theequation}{A\arabic{equation}}
\setcounter{equation}{0}
Given a set of measurement outcomes $x$, an estimator $\hat{\alpha}(x)$ assigns
to each outcome an estimate $\alpha = \hat{\alpha}(x)$ of the unknown parameter $\alpha$. A fundamental result of estimation theory states that the variance of
any unbiased estimator is bounded from below by the CRB,
\begin{equation}
\mathrm{Var}(\hat{\alpha})
\ge
\frac{1}{N\,F(\alpha)},
\label{eq:CRB_classical}
\end{equation}
where $N$ denotes the number of independent measurements and $F(\alpha)$ is the Fisher information. Fisher information has inverse-squared units of the measured parameter and, through the CRB, determines the best resolution achievable from a single measurement. For any quantum measurement, the corresponding Fisher information
is upper bounded by the QFI, leading to the ultimate precision
limit given by the QCRB. The QFI depends only on the quantum states of the statistical model and is independent of the specific measurement performed.

The QFI is commonly interpreted as a measure of how much information about an unknown parameter can, in principle, be extracted from a quantum state. It can be defined via the symmetric logarithmic derivative in the eigenbasis of the 
density operator $\hat{\rho}(\alpha)$ as
\begin{equation}
F_Q^{(S)}(\alpha)
=
2 \sum_{j,k}
\frac{
\left|
\langle \psi_j | (\partial_\alpha \hat{\rho}(\alpha)) | \psi_k \rangle
\right|^2
}{
p_j + p_k
},
\end{equation}
where $p_i$ and $|\psi_i\rangle$ denote the eigenvalues and eigenstates of
$\hat{\rho}(\alpha)$, respectively.

Since both the eigenvalues and eigenstates may depend on the parameter
$\alpha$, the QFI naturally separates into two distinct
contributions, which can be written as
\begin{equation}
F_Q^{(S)}(\alpha)
=
\sum_i
\frac{(\partial_\alpha p_i)^2}{p_i}
+
2 \sum_{j \neq k}
\frac{(p_k - p_j)^2}{p_j + p_k}
\left|
\langle \psi_j | \partial_\alpha \psi_k \rangle
\right|^2 .
\label{FI-def}
\end{equation}

The first term is associated with
variations of the probability distribution $p_i(\alpha)$, while the second
term represents the additional quantum contribution arising from
parameter-dependent eigenstates. If the eigenstates of
$\hat{\rho}(\alpha)$ are independent of $\alpha$, the second term vanishes and
the QFI reduces to the classical-like contribution from the eigenstates. In a limited sense, the classical contribution is sensitive only to amplitudes, whereas the quantum contribution encodes information from both amplitudes and phases.

The QFI satisfies several important properties.
In particular, it is additive for independent quantum systems. For two
independent states $\hat{\rho}_1(\alpha)$ and $\hat{\rho}_2(\alpha)$, the QFI
obeys
\begin{equation}
F_Q^{(S)}\!\left(\hat{\rho}_1(\alpha)\otimes \hat{\rho}_2(\alpha)\right)
=
F_Q^{(S)}\!\left(\hat{\rho}_1(\alpha)\right)
+
F_Q^{(S)}\!\left(\hat{\rho}_2(\alpha)\right).
\label{eq:QFI_additivity}
\end{equation}

Another key property of the QFI is convexity with respect to quantum states.
For two states $\hat{\rho}_1(\alpha)$ and $\hat{\rho}_2(\alpha)$ prepared with
probabilities $p_1$ and $p_2$ such that $p_1 + p_2 = 1$, the QFI satisfies
\begin{equation}
F_Q^{(S)}\!\left(p_1 \hat{\rho}_1(\alpha) + p_2 \hat{\rho}_2(\alpha)\right)
\le
p_1 F_Q^{(S)}\!\left(\hat{\rho}_1(\alpha)\right)
+
p_2 F_Q^{(S)}\!\left(\hat{\rho}_2(\alpha)\right).
\label{eq:QFI_convexity}
\end{equation}

For a pure quantum state $|\psi(\alpha)\rangle$ depending smoothly on the same
parameter $\alpha$, the QFI simplifies to
\begin{equation}
F_Q^{\mathrm{pure}}(\alpha)
=
4\left(
\langle \partial_\alpha\psi|\partial_\alpha\psi\rangle
-
\left|\langle \psi|\partial_\alpha\psi\rangle\right|^2
\right),
\label{eq:pure_QFI}
\end{equation}
where $|\partial_\alpha\psi\rangle \equiv
\frac{\partial}{\partial\alpha}|\psi\rangle$. The above expression is employed for pure-state neutrinos in the subsequent analysis.


\end{document}